\documentclass{emulateapj}
\usepackage{bm}
\usepackage{color}
\usepackage{multirow}
\usepackage{graphicx}
\usepackage{longtable}
\usepackage{rotating}
\usepackage{threeparttable}
\usepackage{epsfig}
\usepackage{amssymb,amsmath}
\shorttitle{The luminosity dependence of quasar UV continuum slope: dust extinction scenario}
\shortauthors{Xie et al.}

\begin{document}

\title{The luminosity dependence of quasar UV continuum slope: dust extinction scenario}

\author{Xiaoyi Xie\altaffilmark{1,2},
Zhengyi Shao\altaffilmark{1,3\dag},
Shiyin Shen\altaffilmark{1,3},
Hui Liu\altaffilmark{1,2},
Linlin Li\altaffilmark{1,2}
}

\affil{\altaffilmark{1} Key Laboratory for Research in Galaxies and Cosmology,
  Shanghai Astronomical Observatory, Chinese Academy of Sciences, 80 Nandan Road, Shanghai 200030, China}
\affil{\altaffilmark{2} Graduate University of the Chinese Academy of Sciences, No.19A Yuquan Road, Beijing 100049, China}
\affil{\altaffilmark{3} Key Lab for Astrophysics, Shanghai 200234, China}
\altaffiltext{\dag}{zyshao@shao.ac.cn}

\begin{abstract}
We investigate the UV continuum slope $\alpha$ of a large quasar sample from SDSS DR7.
By using specific continuum windows, we build two samples at lower ($0.71<z<1.19$) and higher ($1.90<z<3.15$) redshifts,
which correspond to the continuum slopes at longer (NUV) and shorter (FUV) rest wavelength ranges respectively.
Overall, the average continuum slopes are $-0.36$ and $-0.51$ for $\alpha_{\rm NUV}$ and $\alpha_{\rm FUV}$ with similar dispersions $\sigma_{\alpha} \sim 0.5$.
For both samples, we confirm the luminosity dependence of the continuum slope, i.e., fainter quasars have redder spectra.
We further find that both $\alpha_{\rm NUV}$ and $\alpha_{\rm FUV}$ have a common upper limit ($\sim 1/3$) which is almost independent of the quasar luminosity  $L_{\rm bol}$.
This finding implies that the intrinsic quasar continuum (or the bluest quasar), at any luminosity, obey the standard thin disk model.
We propose that the other quasars with redder $\alpha$ are caused by the reddening from the dust {\it locally}.
With this assumption, we employ the dust extinction scenario to model the observed $L_{\rm bol}-\alpha$ relation.
We find that, a typical value of $E(B-V)\sim0.1$ to $0.3$ mag (depending on the types of extinction curve) of the quasar {\it local} dust is enough to explain the discrepancy of $\alpha$ between the observation ($\sim-0.5$) and the standard accretion disk model prediction ($\sim 1/3$).
\end{abstract}

\section{Introduction}\label{sect:intro}
Quasar continuum in the UV-Opt band is powered by the thermal energy of the accretion disk and typically written as a power-law relation $f_\nu \propto \nu^\alpha$, where $\alpha$ is the continuum slope and a larger $\alpha$ means bluer/flatter continuum.
The standard thin disk (STD) model predicts a constant value $\sim1/3$ of the slope in the UV-Opt band \citep{Lynden69,Shakura73}.
However, in observation, the average $\alpha$ of this band is about $-0.5$ \citep{Vanden01}, which is far more redder than the theoretical prediction \citep{Davis07,Shankar16}.
Although more sophisticated disk models do predict redder/softer slopes when they include more details such as the comptonization in the disk atmosphere,
the deviation from local thermodynamic equilibrium and so on, they are still not as red as the observations \citep[e.g.,][]{Hubeny00}.
To alleviate the discrepancy between model prediction and observations, several other explanations, e.g., dust extinction, host galaxy contamination have been proposed \citep{Koratkar99}.

The internal quasar dust (e.g., dust torus), together with the dust in quasar host galaxy, attenuates the UV-Opt spectrum, which prevents us from knowing the intrinsic quasar spectrum.
What's more, the reddening effect may be luminosity dependent.
Inferring from the lack of correlation of Lyman $\alpha$ equivalent width with the quasar UV slope,  \cite{Cheng91} concluded that the intrinsic shape of the UV spectrum is independent of luminosity over a range of three orders of magnitude.
Combining the findings in \cite{Cheng91}, \cite{Gaskell04} further concluded that the dust reddening of the UV continuum is luminosity-dependent in the way that the low luminosity quasars are more reddened.
Such a simple dust extinction model, instead of sophisticated accretion disk model, can nicely explain the observational result that quasar UV continuum is redder toward lower luminosity \citep{Carballo99,Telfer02,Davis07}.

In order to derive and then correct the dust extinction of quasars, a ``pair method'' is often used by comparing a reddened quasar with an unreddened one.
There are many choices of the unreddened quasar spectrum, e.g., the composite quasar spectrum \citep{Czerny04,Gaskell04}, the spectrum of certain blue quasars \citep[hereafter GB07]{Gaskell07}.
However, the dust properties derived from different methods/samples are controversial \citep{Czerny07,Li07}.
To have a better understanding of the dust properties in this extinction model, a systematical study of the UV continuum slopes of a large and well-defined quasar sample is required.

Strictly speaking, all types of quasars, even the type I quasars, suffer some extent of {\it local} dust extinction, either from the accretion disk vicinity or the host galaxy.
\cite{Gaskell15} noticed that even the bluest 10\% of the low redshift AGNs targeted by the SDSS have significant reddening.
Therefore, the choices of composite spectra or blue quasars as the reference for deriving the dust extinction in other reddened quasars still take the risks to suffer from the extinction of {\it local} dust.
In this case, the extinction/reddening will be underestimated systematically.

Alternatively, one may assume that the quasar continuum predicted by the theoretical model is the one with non-dust extinction and then get constraints on the dust properties of  the observed quasars \citep{Davis07}.
In this paper, we test such an idea with the STD model and a large sample of quasars from the Sloan Digital Sky Survey (SDSS) to draw more information of quasar dust.

This Paper is organized as follows.
In Section 2, we introduce the quasar sample, define and measure their continuum slopes.
Then, we describe the slope distributions and their luminosity dependence.
In Section 3, we take the STD model to calculate the intrinsic quasar continuum slope and then make a quantitative comparison with the observational data to evaluate the dust extinction scenario.
We make further discussions on the possible explanations of the luminosity-extinction dependence  in Section 4.
Finally, we draw a brief conclusion in Section 5.
Throughout the paper we assume a concordance $\Lambda$-cosmology with $h = 0.7$, $\Omega_{\rm m} =0.3$ and $\Omega_{\Lambda}=0.7$.

\section{Observational Slopes and their luminosity dependence}\label{sec:data}
\subsection{Quasar sample and continuum slope measurement}\label{sec:sample}
Our sample is defined from the SDSS Data Release 7 (DR7) quasar catalog (105783 objects),
which are selected to be brighter than $M_i=-22.0$ and have at least one emission line with FWHM (full width at half maximum) greater than 1000 km s$^{-1}$ or have interesting/complex absorption features.
In the following measurements of the continuum slopes, we take the residual sky emission lines subtracted spectra from \cite{Wild10} and improved redshift measurements from \cite{Hewett10}.
All these spectra are then corrected for the Galactic reddening by using the SFD map of \cite{Schlegel98} and the extinction curve of \cite{Cardelli89}.
We also take bolometric luminosity ($L_{\rm bol}$), black hole mass ($M_{\rm BH}$), Eddington accretion ratio ($L_{\rm bol}/L_{\rm Edd}$) values and broad absorption line (BAL) flags from \cite{Shen11}.

To avoid the influence of emission and absorption lines in measuring the slope, specific continuum windows are often selected \citep{Vanden01,Davis07,Xie15}.
Following \cite{Xie15}, we take three windows, 1350-1365$\rm\AA$, 2210-2230$\rm\AA$ and 4200-4230$\rm\AA$.
Since the SDSS spectra cover the observed wavelength range from $\sim3800$ to $\sim9200$ $\rm\AA$, therefore, when an individual spectrum is transformed to its rest frame, it covers at most two adjacent windows.
For quasars with $0.71 < z < 1.19$, we can measure their slopes from two longer wavelength windows and label them as $\alpha_{\rm NUV}$.
Similarly, for quasars with $1.81 < z < 3.15$, we can measure their slopes $\alpha_{\rm FUV}$ from two shorter wavelength windows.
In practice, for the FUV sample, we further constrain the redshift range to $1.90 < z < 3.15$ in order to keep consistence with the same definition of $L_{\rm bol}$ estimation in Sec.~\ref{sec:eVector}.

We exclude spectra of:  (1) BAL objects; (2) with no calculated $L_{\rm bol}$ or $M_{\rm BH}$; (3) with too many negative flux values in the covered continuum windows.
After that, we have two samples: 19258 quasars in the NUV sample and 19671 quasars in the FUV sample.
For simplicity, we use a log-linear fit for the fluxes in the two covered windows to measure the continuum slope and its uncertainty.
In fact, the spectrum may not follow an explicit power-law.
Nevertheless, this simple method provides a good approximation of $\alpha$, and still can fairly reflect the difference between the reddened and unreddened (model) spectra that we concern.

\subsection{Slope results and their luminosity dependence}
For the NUV sample, the mean and dispersion values of $\alpha_{\rm NUV}$ are $-0.36$ and $0.46$, and for the FUV sample the respective values are $-0.51$, and $0.48$.
Generally, these mean values agree well with most of the previous works \citep{Cheng91,Vanden01,Gaskell04} with the scatter identical to that of \cite{Cheng91} ($\sim0.46$).
On average, the apparent slope of quasars are much smaller (redder) than the theoretically expected value, while the large dispersions imply the complicate mechanism of shaping the spectrum.

\begin{figure*}[htb]
\hspace*{0.4cm}  \epsscale{1}
\includegraphics[scale=.70,angle=270]{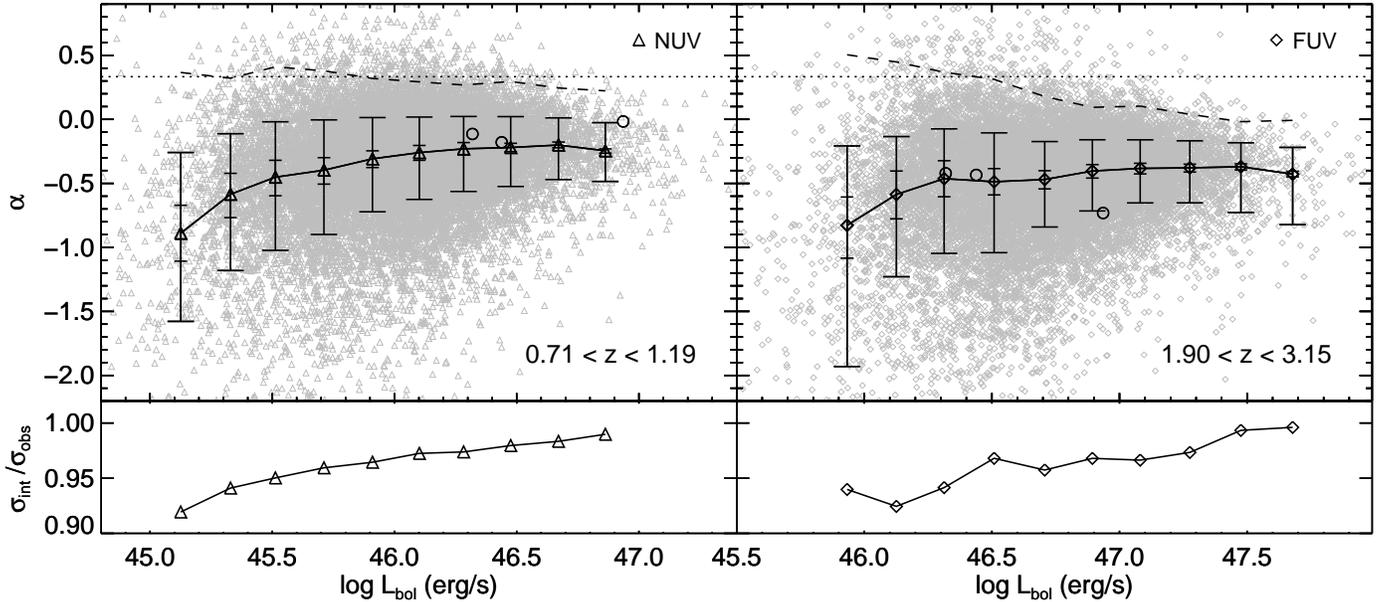}
\caption{The upper panel: the continuum slopes against luminosity for both low redshift (left panel for $\alpha_{\rm NUV}$) and high redshift (right panel for $\alpha_{\rm FUV}$) quasars.
Gray symbols are for individual quasars. The median values of $\alpha$ for each $\log L_{\rm bol}$ bin are linked with solid lines.
The inner error bars represent the median uncertainties of $\alpha$ for each bin,
while the outer error bars are the ranges of 15.9\%-50.0\%th and 50.0\%-84.1\%th percentiles ranges (standing for $\pm \sigma$ of the dispersion).
The 97.7\%th percentiles ($2\sigma$ position) are shown as dashed lines. The horizonal dotted lines locating at the slope value of 1/3 approximate the value of the STD model.
The hollow circles are measurements of $\alpha_{\rm NUV}$ and $\alpha_{\rm FUV}$ for the three bluest spectra in GB07.
The lower panel: the ratios of intrinsic to observational dispersions ($\sigma_{\rm int}/\sigma_{\rm obs}$) of the simulated sample, as a function of the luminosity.}
\label{Fig1}
\end{figure*}

In Fig.~\ref{Fig1}, we plot the continuum slopes  of each quasar against their bolometric luminosity $L_{\rm{bol}}$ in gray symbols.
To check the luminosity dependence of $\alpha$, we further group the quasars into different luminosity bins with a bin size of 0.2 dex and calculate the median values,
dispersions (the outer error bars shown as 15.9\%th to 84.1\%th percentiles of distribution that represent the $\pm \sigma$ ranges).
The median measurement errors of $\alpha$ for each bins are shown as the inner error bars.
Clearly, for both NUV and FUV samples, the median values of the continuum slope increase with luminosity, especially at the low luminosity side, which means less luminous quasars have redder (softer) spectra.
Meanwhile, the dispersions of both $\alpha_{\rm NUV}$ and $\alpha_{\rm FUV}$ are decreasing with luminosity, which means the spread of the $\alpha$ is narrower toward higher luminosity.
These two luminosity dependence confirm to the phenomena that shown in the Fig.4 of \cite{Gaskell04}, though their work only contains a much smaller sample of AGNs.

We notice that the errors of $\alpha$ are also related to the luminosity, with less luminous quasars having larger measurement uncertainties.
As shown in Fig.~\ref{Fig1}, for all luminosity bins, the apparent dispersions of $\alpha$ are significantly larger than the median errors, so they are definitely {\it not} dominated by errors.
However, to estimate the effect of error, we use the Monte-Carlo process which includes 100 sets of artificial samples that mimic our observational samples with similar $L-\alpha$ distributions.
These simulated targets are further randomly modified with their $\alpha$ values according to the possible errors that depend on $L$ and $\alpha$.
Then, we can compare the simulated samples with and without errors of $\alpha$, and find that the changes of the median values of $\alpha$ for all luminosity bins are neglectable,
while the changes of dispersions are all very small with the biggest variation being $\sim10\% $ at the low luminosity side (see also the ratios of intrinsic to observational dispersions $\sigma_{\rm int}/\sigma_{\rm obs}$ for
the simulated sample that plotted in the lower panels of Fig.\ref{Fig1}).
Thus, we claim that the luminosity dependence of continuum slopes and their spreads are subsistent, which reflects some real physical conditions that drive the spectrum shape of AGN.

There is another important thing to note that the continuum slopes of almost all sampled quasars are less than $1/3$ (shown as the dotted horizonal lines in Fig.\ref{Fig1}).
To illustrate this, we plot the dashed lines for the 97.7\%th percentile (standing for the $2\sigma$ level).
They also curve the upper/blue envelope of the slopes.
This upper limitation is roughly luminosity independent except for a little increasing at the low luminosity side of $\alpha_{\rm FUV}$.
It could be partly attributed to their larger observational errors of $\alpha$, thus more quasars will probably have the $\alpha$  measurements over the blue limitation, though they are intrinsically below it.

There are previous works that claim the luminosity  independent cutoff of $\alpha$ at the blue side, i.e. the Fig.4 of \cite{Gaskell04}.
What we want to emphasize here is, based on the large quasar sample from the SDSS, this upper limitation is close to the STD model prediction, more or less.
On the other hand, we understand that $\alpha \sim 1/3$ is only a rough approximation of the model prediction of the intrinsic slope.
Actually, it should be slightly modified by the central black hole mass and the accretion rate (see Sec.\ref{sec:ModelSlope} for details).
Anyway, the distribution of both $\alpha_{\rm NUV}$ and $\alpha_{\rm FUV}$ give us a strong hint that the blue envelope of $\alpha$ is highly possible to exist, while those bluest ones are bare quasars with little (or
no) dust extinction so that their continuum slope are in good consistence with the STD model.

\section{Model and Algorithm}\label{sec:model}
In this section, we assume a simple scenario that the intrinsic continuum radiation of quasars obeys the STD model, then be dimmed and reddened by {\it local} dust,
either from the optical thin dust in the internal region or from the interstellar dust in the host galaxy.
Then, we take the discrepancy between observational slopes and model predictions, either for $\alpha_{\rm NUV}$ or for $\alpha_{\rm FUV}$,
to quantify the dust extinction, as well as to transform the apparent $L-\alpha$ trend to the essential $L$-extinction relation.

\subsection{Intrinsic continuum slope}\label{sec:ModelSlope}
We take the radiatively efficient, geometrically thin and optically thick accretion disc or the so-called standard thin disk (STD) to model the quasar continuum emission,
where the radiation at each radius in the accretion disk is a blackbody \citep{Shakura73}.
Following \cite{Li08}, we adopt the temperature distribution as function of radius according to \cite{Kato98},
\begin{equation}\label{T}
    T(r)=6.9\times10^{7}\alpha_{\rm{ss}}^{-1/5}m^{-1/5}\dot{m}^{3/10}r^{-3/4}f^{3/10} \rm{K},
\end{equation}
where $\alpha_{\rm ss}=0.1$ is the viscosity parameter, $m=M_{\rm BH}/M_{\sun}$ is the black hole mass, $\dot{m}=\dot{M}/\dot{M}_{\rm Edd}=L_{\rm bol}/L_{\rm Edd}$ is the accretion rate in unit of Eddington
accretion rate with $\dot{M}_{\rm Edd}=1.5\times10^{18}{\rm m~g ~s}^{-1}$, $r=R/R_{\rm s}$ ($R_{\rm s}=2{\rm G}M_{\rm BH}/{\rm c}^2$) and $f\equiv 1-(3/r)^{1/2}$.
The shape of the spectrum is then calculated with
\begin{equation}\label{fnu}
    f_\nu \propto \nu^3 \int_{r_{\rm in}}^{r_{\rm out}} \frac{r\,dr}{\exp ({\rm h}\nu/{\rm k}T(r))-1} \quad,
\end{equation}
where $r_{\rm in}=3R_{\rm s}$, $r_{\rm out}$ is the outer radius of disk and is set to be the place where $T(r)$ deceases to 1000K,
$\rm h$ is the Plank constant, $\rm k$ is the Boltzmann constant and $\rm c$ is the speed of light.

\begin{figure}[h]
\hspace*{-1.cm}  \epsscale{1}
\includegraphics[scale=.4,angle=270]{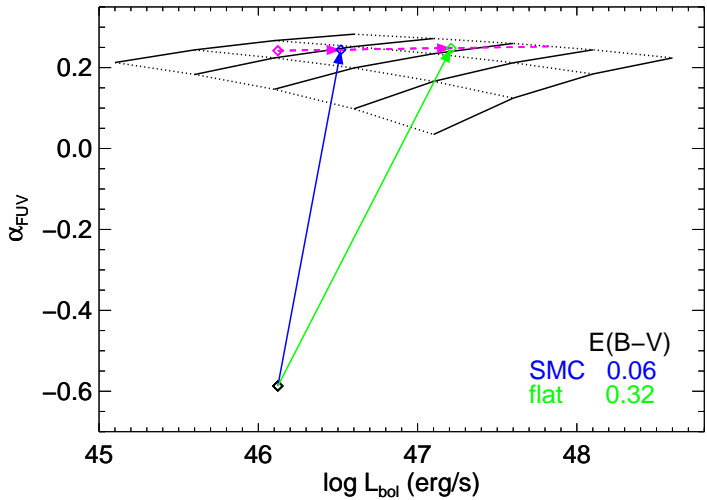}
\caption{Model predicted slopes and dust extinction vectors.
Black solid lines from left to right have $\log M_{\rm BH}$ from 8.5 to 10.5 evenly spaced.
Dotted lines from bottom to top have $\log \dot m$ from -1.5 to 0.0 evenly spaced.
The black diamond is an observed data point.
The magenta dashed line is the trail of the luminosity {\it vs} model continuum slope, while the blue and green arrows stand for the luminosity {\it vs} observed continuum slope trails,
and their intersections represent the intrinsic luminosities and continuum slopes.
Values in the right corner show the extinction needed for the SMC extinction curve and the flat curve (GB07) respectively.}
\label{Fig2}
\end{figure}

In Fig.~\ref{Fig2}, we plot the model slope $\alpha_{\rm FUV}$, as a function of different black hole masses and Eddington accretion rates.
It shows that $\alpha_{\rm FUV}$ decreases with $\log M_{\rm BH}$ and increases with $\log (L_{\rm bol}/L_{\rm Edd})$.
The grid of $\alpha_{\rm NUV}$ is similar to the $\alpha_{\rm FUV}$, with slightly larger slope values and narrower range.
Generally, both of them are close to the value of 1/3, which confirms the expectation from the STD model.

\subsection{Dust extinction vector}\label{sec:eVector}
With the {\it local} dust extinction, the quasar luminosity is dimmed and its continuum spectrum is reshaped (reddened).
If we simplify the intrinsic spectrum as $f_\lambda\propto\lambda^{-2-\alpha}$, then the reddened spectrum becomes $f_\lambda e^{-\tau_\lambda}$,
where the optical depth $\tau_\lambda$ equals to $el_\lambda \cdot E(B-V)/1.086$ with $el_\lambda=A_\lambda/E(B-V)$ refers to the extinction curve.

The reddening of the continuum slope can be derived by
\begin{equation}\label{da}
    \Delta \alpha =  [0.4(el_{\lambda_1}-el_{\lambda_2})\log (\lambda_1/\lambda_2)] \cdot E(B-V)
\end{equation}
where subscript 1 and 2 stand for the selected continuum windows.

According to \cite{Shen11}, $L_{\rm bol}$ is calculated from a monochromatic continuum luminosity and its bolometric correction ($BC$) on the basis of redshift,
\begin{equation}\label{Lb}
    L_{\rm{bol}}=
\left\{
\begin{aligned}
  &  BC_{5100\rm\AA}\times 5100L_{5100{\rm\AA}},\, (z <0.7) \\
  &  BC_{3000\rm\AA}\times 3000L_{3000{\rm\AA}},\, (0.7 \leq z < 1.9) \\
  &  BC_{1350\rm\AA}\times 1350L_{1350{\rm\AA}},\, (z \geq 1.9).
\end{aligned}
\right.
\end{equation}
So, the dimming of $L_{\rm bol}$ can be expressed as the change of corresponding continuum luminosity,
\begin{equation}\label{dL}
    \Delta \log L_{\rm{bol}}=  \Delta \log L_{\lambda} =-0.4el_{\lambda}\cdot E(B-V)
\end{equation}
where $\lambda=3000\rm\AA$ for NUV or $1350\rm\AA$ for FUV samples.

We can regard $(\Delta \log L_{\rm{bol}}, \Delta \alpha)$ as $(x,y)$ components of the ``extinction vector'' ($\vec{e}$).
Obviously, for given continuum windows, the direction of $\vec{e}$ only depends on $el_\lambda$, and its length is in proportional to $E(B-V)$.

Here, we test two typical dust extinction curves, the SMC curve \citep{Gordon03} and the flat AGN curve (GB07).
For the SMC curve, the unit vector of $\vec{e}$ (with $E(B-V)=1$) is $(-2.17,-6.33)$ for NUV sample and $(-6.16, -12.93)$ for FUV sample.
Correspondingly, in the case of the flat curve, these two vectors are $(-2.34,-3.94)$ and $(-3.39, -2.60)$.
We notice that the SMC curve is much steeper in the UV band, so it needs much smaller $E(B-V)$ for a given $\Delta \alpha$.

\subsection{Estimation of $E(B-V)$}\label{sec:ebv}
In Fig.~\ref{Fig2}, as an example, we illustrate the procedure of dust correction for a ``typical" quasar, with observed value of $\log L_{\rm bol} = 46.1$, $\alpha_{\rm FUV}=-0.59$ and $\log M_{\rm BH}=8.84$
(the second point from left in the upper right panel of Fig.~\ref{Fig1}).
For a given extinction curve, the corrections of $L_{\rm bol}$ and $\alpha$ start from the observed value (the black diamond) and follow the opposite direction of the ``extinction vector",
until it reaches its intrinsic $L_{\rm bol}$ and $\alpha$.

On the other hand, the estimation of virial mass of BH is based on both broad emission line width and continuum luminosity.
Since the width of emission line is not affected by reddening, according to the Eq.~2 of \cite{Shen11}, we can easily find that the dust extinction correction on  the black hole mass estimation is
\begin{equation}\label{dM}
    \Delta \log M_{\rm BH} = b~\Delta \log L_{\lambda} = b~\Delta \log L_{\rm bol},
\end{equation}
where $L_\lambda$ is the corresponding continuum luminosity of the broad line and $b$ is the empirical coefficient and equals to 0.62 and 0.53 for NUV and FUV samples respectively \citep{Shen11}.

For each observed quasar, when $L_{\rm bol}$ and ${\rm log} M_{\rm BH}$ are derived, the accretion rate ${\rm log} \dot{m}$ is also fixed.
Therefore, after a dust correction of $E(B-V)$, we can also get an estimation of the intrinsic continuum slope from the STD model presented in Section \ref{sec:ModelSlope}.
Such a correction is displayed by the trace of the magenta dashed line in Fig.~\ref{Fig2}.
When the dust extinction in the quasar is properly accounted for, the corrections from the extinction curve and theoretical model will lead to the consistent result of $\alpha$.
In other words, the intersection point of the two above-mentioned traces represents the intrinsic properties of the ``example'' quasar.
Thus, we can use this point to calculate the $E(B-V)$ that we concern.

It should be noticed that, for our current sample, the observed $M_{\rm BH}$, $L_{\rm bol}$ and $\alpha$ of individual quasars have large uncertainties.
Thus, it seems to be too complicate and uncertain to extract the dust extinction for individual quasars.
Alternatively, we use the median values of the 20 binned sub-samples in Fig.~\ref{Fig1} to be the ``median quasars" that represent their typical features.
So in the following analysis, we only focus on these 20 ``median quasars" and quote their median values of $\alpha_{\rm NUV}$, $\alpha_{\rm FUV}$, $\log L_{\rm bol}$ and $\log M_{\rm BH}$ respectively.

\subsection{$L-E(B-V)$ relation}\label{sec:L-E}
\begin{figure}[h]
\hspace*{-1.cm}  \epsscale{1}
\includegraphics[scale=0.4,angle=270]{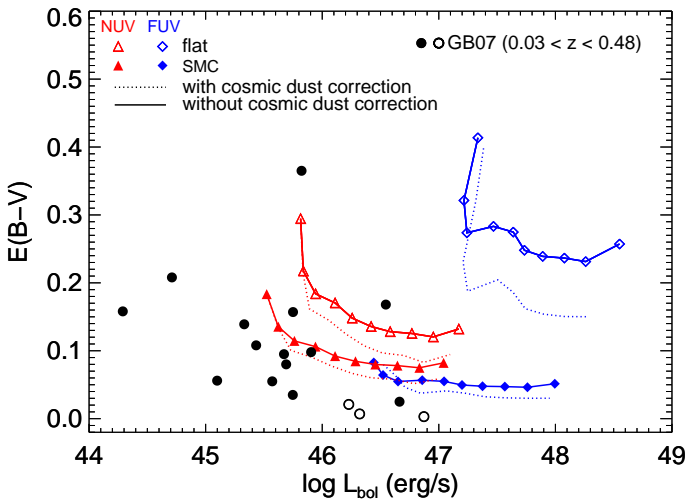}
\caption{The $L-E(B-V)$ relations.
Red triangles are for the  $\alpha_{\rm NUV}$ sample and blue diamonds are for $\alpha_{\rm FUV}$, with solid symbols for the SMC type dust and hollow ones for the flat dust curves.
Dotted lines show results after the cosmic dust correction (Sec.~\ref{sec:CosmicDust}).
Solid and hollow circles (the three bluest ones) are the results of individual quasars for GB07.}
\label{Fig3}
\end{figure}

By applying the algorithm of Sec.\ref{sec:ebv}, we calculate $E(B-V)$ and intrinsic $L_{\rm bol}$ for all 20 ``median quasars'', for both SMC and flat extinction curve situations, and plot the results in Fig.~\ref{Fig3}.
Intuitively, one can conclude that: (1) there is a significant $L-E(B-V)$ anti-correlation with lower luminosity quasars having more dust extinction, no matter which sample (NUV or FUV) or which extinction curve we use;
(2) for the SMC curve, the required $E(B-V)$ value is only about $0.05 \sim 0.2$, while for the flat curve it increases to $0.15 \sim 0.4$.
The systematical higher $E(B-V)$ values for the flat extinction curve is simply because that the SMC extinction curve is much steeper in the UV band, which requires smaller $E(B-V)$ to produce the same reddening of $\alpha$.

For comparison, we also plot the results  from GB07 in Fig.~\ref{Fig3}, which used the ``pair method'' and the flat extinction curve.
These isolated points disperse largely but have the same $L-E(B-V)$ trend as ours.
Compared with the same extinction curve, these points are mostly below our results.
This result is comprehensible.
The quasar sample of GB07 comes from \cite{Shang05} which is chosen to be bright in the far-UV, thus it might be biased toward low-reddening objects.
More importantly, the bluest quasars (PG0953+414, PG1100+772, 3C273) they choose as the unreddened spectra may still contain a certain extent of dust extinction.
We have measured the $\alpha_{\rm NUV}$ and $\alpha_{\rm FUV}$ of these three quasars and find that they are obviously redder than the theoretical values of the bare AGNs (see the hollow circles in Fig.~\ref{Fig1}).

Comparing the quasars with the same intrinsic $L_{\rm bol}$ but in different samples (NUV and FUV), we see that the $E(B-V)$ values derived from the SMC extinction curve  are more consistent.
However, this apparent consistence does not motivate us to favor the SMC type extinction curve in quasars.
Because the $E(B-V)$ values we derived are strongly dependent on the intrinsic slopes of the quasar continuum, which is far from well-defined in the simple STD model we adopted.
If the real intrinsic FUV slope of the quasars are softer than our model calculation, then our derived $E(B-V)$ would be certainly biased to higher values for the flat extinction curve while change less for the SMC curve.
Besides, the NUV and FUV samples are actually in different redshifts and surely have the observational bias with higher redshift sample (FUV) containing more luminous quasars.
Therefore the direct comparison of their $E(B-V)$ values maybe unfair.
Moreover, given the FUV sample at higher redshift, it is expected that these quasars are more attenuated considering the cosmic dust along the line of sight \citep{Xie15} (see more discussions in Sec.~\ref{sec:CosmicDust}).

\section{Discussion}\label{sec:Discussion}
In this study, we introduce a luminosity dependent dust extinction model and successfully apply it to explain the phenomena that the more luminous quasars have bluer continuum.
In this section, we further discuss its physical implications and other possibilities.

\subsection{luminosity dependence of dust extinction}
Firstly, the relative bluer continuum slope of more luminous quasar is unlikely to be solely caused by the intrinsic properties of the accretion disk.
The more sophisticated model are also hard to produce such a red spectrum \citep{Davis07}.
Besides, a softer continuum at lower luminosity can hardly explain their larger broad emission line equivalent widths,
i.e. the well-known Baldwin effect \citep{Baldwin77}.

On the other hand, it is quite straightforward to ascribe the observed $L-\alpha$ relationship to the quasar {\it{local}} dust extinction.
The {\it{local}} dust extinction may physically be originated from the internal dust around the accretion disk, or from the dust in the host galaxy.
For the quasar host galaxies, observations show that their star formations are co-evolved with the growth of the black holes \citep{Gebhardt00,Ferrarese00,Kormendy13}.
Therefore, we would expect that there are more star formation, so that more dust generated in the host galaxies of high luminosity quasars.
Such a scenario is obviously not consistent with the trend we required to explain the observed $L-\alpha$ relation.
Therefore, the {\it local} dust in our model is more likely to be dominated by the internal  one.
If so, it is worthwhile to further discuss whether the phenomenological internal dust in our model is compatible with the physical dust torus model in the revised unified AGN paradigm \citep{Netzer15}.

In this revised model, the dust tori are packed with gas/dust into clumps, and the intrinsic attenuation of an AGN is determined by the number of the clumps along the line-of-sight,
their optical depth and distribution \citep[e.g.][]{Nenkova08,Honig13}.
Since our sample of quasars are type I sources, the photons from the accretion disk have only crossed few optical-thin clouds.
These clouds are probably originated from the transition area between the dust torus (optical thick) and broad-line-region (BLR, dust-free) or the low density interclump medium \citep{Nenkova08,Li15}
and provide the attenuation of the type I AGNs.
From the IR reverberation mapping observation, we know that the dust sublimation radius $R_{\rm sub}$ is proportional to the quasar luminosity by $R_{\rm sub} \propto L^{1/2}$ \citep{Suganuma06}.
Thus the distribution of the internal dust will depend on the luminosity of the central engine in a statistical way that the dust tori of higher $L_{\rm bol}$ sources have smaller effective covering factors \citep{Netzer15}.
Such a scenario naturally explains the higher number ratio of type I/type II sources and lower $L_{\rm IR}/L_{\rm bol}$ ratios in higher $L_{\rm bol}$ quasars \citep{Simpson05,Shen10,Gu13,Toba14}.
If we assume the sizes of these optical thin clouds are smaller (so that with smaller optical depths) for higher $L_{\rm bol}$ AGNs due to the stronger dust sublimation effect,
we would expect that even for type I configuration, the higher $L_{\rm bol}$ AGNs are averagely less attenuated than their lower luminosity counterparts.

Apart from the distribution of dust clumps, the dust properties or in terms of the extinction curve may be another factor which leads to the different reddening in high and low luminosity quasars.
From the theoretical consideration, the larger dust grains are preferred to stay in the severe radiation conditions near the nuclei as the small grains are more easy to be destroyed or blown away by strong radiation \citep{Gaskell04}.
Therefore, in the case of low luminosity quasars, the dust component is biased to small dust grains and more UV photons will be absorbed, leading to smaller observed $\alpha$, which is consistent with our finding.

\subsection{redshift dependence of $\alpha$}\label{sec:CosmicDust}
 In a recent work, we have found that, within the same luminosity, quasar continuum becomes redder toward higher redshift, which is successfully attributed to the accumulation effect of the cosmic dust extinction \citep{Xie15}.
Therefore, the reddening from the cosmic dust accumulated in the path should be corrected.

We take the cosmic dust extinction model from \cite{Xie15} (with $n\sigma_{\nu} = 10^{-5} {\rm h~ Mpc^{-1}}$ and the flat extinction curve,
as shown in their Fig.5) and make corrections on the observed $L_{\rm bol}$, $M_{\rm BH}$, $\dot{m}$ and $\alpha$ for each quasar.
After the correction of cosmic dust extinction, we repeat the analysis of the {\it local} dust extinction in Sec.\ref{sec:ebv}.
The new results are plotted as dotted lines in Fig.~\ref{Fig3} for illustration.
As we can see, the involvement of the cosmic dust extinction only slightly reduces the {\it local} dust extinction and does not change the trend of the $L-E(B-V)$ relation.
That is to say, comparing with the extinction from the {\it local} dust, the accumulated dust extinction along the line of sight is a small quantity.

Moreover, we had supposed that the cosmic dust extinction is one of the reasons for the inconsistence of the $L-E(B-V)$ relation of NUV and FUV samples in the case of the flat extinction curve (Sec.\ref{sec:L-E}).
Now we can learn from dotted lines in Fig.\ref{Fig3} to recognize that this factor only slightly alleviates the inconsistence.

\section{Summary}\label{sec:Summary}
In this paper, we investigate the UV continuum slope of a large sample of quasars selected from SDSS DR7.
It is found that for both $\alpha_{\rm NUV}$ and $\alpha_{\rm FUV}$, the observed continuum slope are averagely and significantly redder than theoretical expected values.
The observed $\alpha_{\rm NUV}$ and $\alpha_{\rm FUV}$ show dependence on the quasar luminosity, with higher luminosity quasars having bluer continuum.
Such findings have been reported in many previous studies.
Besides, we notice that the quasar continuum slopes show an upper limit ($\alpha \sim 1/3$) either for $\alpha_{\rm NUV}$ or $\alpha_{\rm FUV}$ at any luminosity range.
This strongly implies intrinsic quasars (or perhaps the observed bluest quasars) obey the standard thin disk model.

We further test the assumption that all observed quasar spectra intrinsically follow the standard model prediction, but get dimming and reddening by the dust surrounding the accretion disk.
We calculate the $E(B-V)$ needed to achieve the intrinsic value of $\alpha$ by simultaneously considering the correction of the bolometric luminosity and the continuum slope due to dust extinction and reddening.
Our estimation of $E(B-V)$ under the flat AGN extinction curve case is consistent with the previous finding $E(B-V)\sim0.3$ \citep{Gaskell15}.
For the SMC type extinction curve, because of its much steeper shape in UV wavelength, the $E(B-V)$ values required is even smaller $\sim 0.1$.

Our results also show that there is a dependence of $E(B-V)$ on intrinsic quasar luminosity $L_{\rm bol}$.
The higher $L_{\rm bol}$ quasars are averagely less reddened than the lower luminosity ones.
We have discussed that such a luminosity dependence is compatible with the revised unified AGN model.
We argue that the type I quasars are slightly attenuated by the optical thin clouds in the transition area where the dust particles are sublimated by the central engine or the interclump medium.
We have also checked the effect of the cosmic dust extinction and conclude that compared with the {\it local} dust extinction, the accumulated dust extinction along the line of sight is a small quantity.

\acknowledgments
We gratefully acknowledge the referee Dr. Martin Gaskell for  very helpful comments and detailed suggestions.
The authors thank Prof. Minfeng Gu and Dr. Shuangliang Li in SHAO for helpful discussions.
This work was supported by the Strategic Priority Research Program ``The Emergence of Cosmological Structures'' of the Chinese Academy of Sciences (CAS; grant XDB09030200),
the National Natural Science Foundation of China (NSFC) with the Project Numbers 11390373, 11433003, and the ``973 Program'' 2014 CB845705.


\begin{thebibliography}{}
\bibitem[Baldwin(1977)]{Baldwin77} Baldwin, J.~A.\ 1977, \apj, 214, 679
\bibitem[Carballo et al.(1999)]{Carballo99} Carballo, R., Gonz{\'a}lez-Serrano, J.~I., Benn, C.~R., S{\'a}nchez, S.~F., \& Vigotti, M.\ 1999, \mnras, 306, 137
\bibitem[Cardelli et al.(1989)]{Cardelli89} Cardelli, J.~A., Clayton, G.~C., \& Mathis, J.~S.\ 1989, \apj, 345, 245
\bibitem[Cheng et al.(1991)]{Cheng91} Cheng, F.~H., Gaskell, C.~M., \& Koratkar, A.~P.\ 1991, \apj, 370, 487
 \bibitem[Czerny et al.(2004)]{Czerny04} Czerny, B., Li, J., Loska, Z., \& Szczerba, R.\ 2004, \mnras, 348, L54
\bibitem[Czerny(2007)]{Czerny07} Czerny, B.\ 2007, The Central Engine of Active Galactic Nuclei, 373, 586
\bibitem[Davis et al.(2007)]{Davis07} Davis, S.~W., Woo, J.-H., \& Blaes, O.~M.\ 2007, \apj, 668, 682
\bibitem[Ferrarese \& Merritt(2000)]{Ferrarese00} Ferrarese, L., \& Merritt, D.\ 2000, \apjl, 539, L9
\bibitem[Gaskell et al.(2004)]{Gaskell04} Gaskell, C.~M., Goosmann, R.~W., Antonucci, R.~R.~J., \& Whysong, D.~H.\ 2004, \apj, 616, 147
\bibitem[Gaskell \& Benker(2007)]{Gaskell07} Gaskell, C.~M., \& Benker, A.~J.\ 2007, arXiv:0711.1013 (GB07)
\bibitem[Gaskell(2015)]{Gaskell15} Gaskell, C.~M.\ 2015, arXiv:1512.09291
\bibitem[Gebhardt et al.(2000)]{Gebhardt00} Gebhardt, K., Bender, R., Bower, G., et al.\ 2000, \apjl, 539, L13
\bibitem[Gordon et al.(2003)]{Gordon03} Gordon, K.~D., Clayton, G.~C., Misselt, K.~A., Landolt, A.~U., \& Wolff, M.~J.\ 2003, \apj, 594, 279
\bibitem[Gu(2013)]{Gu13} Gu, M.\ 2013, \apj, 773, 176
\bibitem[Hewett \& Wild(2010)]{Hewett10} Hewett, P.~C., \& Wild, V.\ 2010, \mnras, 405, 2302
\bibitem[H\"{o}nig(2013)]{Honig13} H\"{o}nig, S.~F.\ 2013, arXiv:1301.1349
\bibitem[Hubeny et al.(2000)]{Hubeny00} Hubeny, I., Agol, E., Blaes, O., \& Krolik, J.~H.\ 2000, \apj, 533, 710
\bibitem[Kato et al.(1998)]{Kato98} Kato, S., Fukue, J., \& Mineshige, S.\ 1998, Black-hole accretion disks.~ Edited by Shoji Kato, Jun Fukue, and Sin Mineshige.~ Publisher: Kyoto, Japan: Kyoto University Press, 1998.~ISBN: 4876980535,
\bibitem[Koratkar \& Blaes(1999)]{Koratkar99} Koratkar, A., \& Blaes, O.\ 1999, \pasp, 111, 1
\bibitem[Kormendy \& Ho(2013)]{Kormendy13} Kormendy, J., \& Ho, L.~C.\ 2013, \araa, 51, 511
\bibitem[Li(2007)]{Li07} Li, A.\ 2007, The Central Engine of Active Galactic Nuclei, 373, 561
\bibitem[Li \& Cao(2008)]{Li08} Li, S.-L., \& Cao, X.\ 2008, \mnras, 387, L41
\bibitem[Li et al.(2015)]{Li15} Li, Z., Zhou, H., Hao, L., et al.\ 2015, \apj, 812, 99
\bibitem[Lynden-Bell(1969)]{Lynden69} Lynden-Bell, D.\ 1969, \nat, 223, 690
\bibitem[Nenkova et al.(2008)]{Nenkova08} Nenkova, M., Sirocky, M.~M., Nikutta, R., Ivezi{\'c}, {\v Z}., \& Elitzur, M.\ 2008, \apj, 685, 160
\bibitem[Netzer(2015)]{Netzer15} Netzer, H.\ 2015, \araa, 53, 365
\bibitem[Schlegel et al.(1998)]{Schlegel98} Schlegel, D.~J., Finkbeiner, D.~P., \& Davis, M.\ 1998, \apj, 500, 525
\bibitem[Schneider et al.(2010)]{Schneider10} Schneider, D.~P., Richards, G.~T., Hall, P.~B., et al.\ 2010, \aj, 139, 2360
\bibitem[Shakura \& Sunyaev(1973)]{Shakura73} Shakura, N.~I., \& Sunyaev, R.~A.\ 1973, \aap, 24, 337
\bibitem[Shankar et al.(2016)]{Shankar16} Shankar, F., Calderone,G., Knigge, C., et al.\ 2016, \apjl, 818, L1
\bibitem[Shang et al.(2005)]{Shang05} Shang, Z., Brotherton, M.~S., Green, R.~F., et al.\ 2005, \apj, 619, 41
\bibitem[Shen et al.(2010)]{Shen10} Shen, S., Shao, Z., \& Gu, M.\ 2010, \apjl, 725, L210
\bibitem[Shen et al.(2011)]{Shen11} Shen, Y., Richards, G.~T., Strauss, M.~A., et al.\ 2011, \apjs, 194, 45
\bibitem[Simpson(2005)]{Simpson05} Simpson, C.\ 2005, \mnras, 360, 565
\bibitem[Suganuma et al.(2006)]{Suganuma06} Suganuma, M., Yoshii, Y., Kobayashi, Y., et al.\ 2006, \apj, 639, 46
\bibitem[Telfer et al.(2002)]{Telfer02} Telfer, R.~C., Zheng, W., Kriss, G.~A., \& Davidsen, A.~F.\ 2002, \apj, 565, 773
\bibitem[Toba et al.(2014)]{Toba14} Toba, Y., Oyabu, S.,Matsuhara, H., et al.\ 2014, \apj, 788, 45
\bibitem[Vanden Berk et al.(2001)]{Vanden01} Vanden Berk, D.~E., Richards, G.~T., Bauer, A., et al.\ 2001, \aj, 122, 549
\bibitem[Wild \& Hewett(2010)]{Wild10} Wild, V., \& Hewett, P.~C.\ 2010, arXiv:1010.2500
\bibitem[Xie et al.(2015)]{Xie15} Xie, X., Shen, S., Shao, Z., \& Yin, J.\ 2015, \apjl, 802, L16

\end{thebibliography}
\end{document}